\RequirePackage[2020-02-02]{latexrelease}
\documentclass[onecolumn,amsmath,amssymb,10pt,aps]{revtex4}
  
\usepackage[utf8]{inputenc}
\usepackage[T1]{fontenc}

\usepackage{amsmath}
\usepackage{amsfonts}
\usepackage{graphicx}
\usepackage{yfonts}
\usepackage{color}
\usepackage[normalem]{ulem}
\usepackage{amsthm}
\usepackage{bm}
\usepackage{bbm}
\usepackage{mathtools}
\usepackage{array}
\usepackage{placeins}
\usepackage{enumitem}

\newcommand\bigforall{\mbox{\Large $\mathsurround=1pt\forall$}}

\def\<{\langle}
\def\>{\rangle}

\newcommand{\Tr}{\mathrm{Tr}}
\def\oper{{\mathchoice{\rm 1\mskip-4mu l}{\rm 1\mskip-4mu l}
{\rm 1\mskip-4.5mu l}{\rm 1\mskip-5mu l}}}
\DeclareMathAlphabet\mathbfcal{OMS}{cmsy}{b}{n}

\mathchardef\mhyphen="2D 

\newtheorem{Definition}{Definition}
\newtheorem{Remark}{Remark}
\newtheorem{Proposition}{Proposition}
\newtheorem{Example}{Example}

\begin{document}


\title{How much symmetry do symmetric measurements need\\for efficient operational applications?}

\author{Katarzyna Siudzi\'{n}ska}
\affiliation{Institute of Physics, Faculty of Physics, Astronomy and Informatics, Nicolaus Copernicus University in Toru\'{n}, ul.~Grudzi\k{a}dzka 5, 87--100 Toru\'{n}, Poland}

\begin{abstract}
We introduce a generalization of symmetric measurements to collections of unequinumerous positive, operator-valued measures (POVMs). For informationally complete sets, we propose construction methods from orthonormal Hermitian operator bases. The correspondence between operator bases and measurements can be as high as four-to-four, with a one-to-one correspondence following only under additional assumptions. Importantly, it turns out that some of the symmetry properties, lost in the process of generalization, can be recovered without fixing the same number of elements for all POVMs. In particular, for a wide class of unequinumerous symmetric measurements that are conical 2-designs, we derive the index of coincidence, entropic uncertainty relations, and separability criteria for bipartite quantum states.
\end{abstract}

\flushbottom

\maketitle

\thispagestyle{empty}

\section{Introduction}

Quantum measurements are important tools widely used in many quantum information processing tasks. Their applications include quantum communication protocols \cite{Zhou,Song}, quantum filtering \cite{Hirsch}, entanglement detection \cite{Blume}, quantum teleportation \cite{Siendong}, and quantum tomography \cite{Prugovecki}. In quantum information theory, measurements are represented via positive operator-valued measures (POVMs), which are sets of positive operators that sum up to the identity operator. Examples of popular POVMs are symmetric, informationally complete (SIC) POVMs \cite{Renes} and mutually unbiased bases (MUBs) \cite{Schwinger,Szarek}. Both SIC POVMs and MUBs are projective measurements because their elements are rank-1 projectors. Their non-projective counterparts -- general SIC POVMs and mutually unbiased measurements (MUMs) \cite{Kalev,Gour} -- find uses in quantum state discrimination and other postprocessing tasks \cite{Ivanovic,Dieks}.

Recently, there has been an increased interest in generalizations of SIC POVMs. Semi-SIC POVMs relax the condition for measurement operators to be of equal trace \cite{semi-SIC}. This simple change leads to discrete values of the Hilbert-Schmidt product between different measurement operators in any $d\geq 3$. An analogical property is observed in equioverlapping measurements, which additionally drop the condition of informational completeness \cite{EOM22,EOM24,EOMq3}. Importantly, both semi-SIC POVMs and equioverlapping measurements are composed of rank-1 projectors.

On a different note, there is a generalization of both SIC POVMs and MUBs that allows for non-projective measurements. Symmetric measurements, or $(N,M)$-POVMs, are collections of $N$ mutually unbiased POVMs, each with $M$ elements \cite{SIC-MUB}. Despite being a relatively recent construction, symmetric measurements have already found applications in characterizations of entropic uncertainty relations \cite{SICMUB_entropic,SICMUB_entropic2}, average coherences \cite{SICMUB_App3}, quantum steerability \cite{Alber2}, and the Brukner-Zeilinger invariants \cite{SICMUB_BZ}. Other implementations of $(N,M)$-POVMs include optimal state estimation \cite{SICMUB_design} and entanglement detection via improved separability criteria \cite{SICMUB_App,SICMUB_App2,Lai,SICMUB_App4,Alber} or entanglement witnesses \cite{SICMUB_Pmaps}.

In this paper, we provide a further generalization of $(N,M)$-POVMs by relaxing the requirement for equal number of POVM elements. In other words, we introduce collections of $N$ POVMs $\mathcal{E}_\alpha$, where each consists in $M_\alpha$ measurement operators. This simple change greatly increases the number of characteristic parameters and complexity, especially for high values of $N$. Now, we pose two important questions: $(i)$ How much symmetry can be recovered without assuming that $M_\alpha=M_\beta$? $(ii)$ Which properties and applications carry over to his generalization?

We start from a definition of generalized symmetric measurements in any finite dimension. We show that they form an informationally complete set if and only if the total number of measurement operators is $d^2+N-1$. Next, we propose four methods of construction from Hermitian orthonormal bases and analyze when these methods give rise to distinct collections of measurement operators. A separate section is dedicated to symmetrizations of generalized symmetric measurements that do not restrict the numbers of POVM elements. Of particular interest if a wide class of unequinumerous symmetric measurements that is a conical 2-design. As it turns out, it possesses enough symmetry properties to allow for analytical calculations of the index of coincidence, entropic uncertainty relations, and separability criteria for bipartite quantum states.

\section{Generalized symmetric measurements}

Recently, we have introduced the notion of a symmetric measurement $\{E_{\alpha,k};\,k=1,\ldots,M;\,\alpha=1,\ldots,N\}$ called an $(N,M)$-POVM. From definition, it is a collection of $N$ POVMs with $M$ elements each \cite{SIC-MUB}. Its distinguishing property is that the measurement operators $E_{\alpha,k}$ satisfy strong symmetry conditions
\begin{equation}\label{M}
\begin{split}
\Tr (E_{\alpha,k})&=w,\\
\Tr (E_{\alpha,k}^2)&=x,\\
\Tr (E_{\alpha,k}E_{\alpha,\ell})&=y,\qquad \ell\neq k,\\
\Tr (E_{\alpha,k}E_{\beta,\ell})&=z,\qquad \beta\neq\alpha,
\end{split}
\end{equation}
where
\begin{equation}\label{xyz}
w=\frac dM,\qquad y=\frac{d-Mx}{M(M-1)},\qquad z=\frac{d}{M^2},
\end{equation}
and $x$ is a free parameter from the range
\begin{equation}\label{x}
\frac{d}{M^2}<x\leq\min\left\{\frac{d^2}{M^2},\frac{d}{M}\right\}.
\end{equation}
Projective measurements follow for $x=d^2/M^2$, which lies in the admissible range only for $M\geq d$. It has been shown that there exist at least four informationally complete families of $(N,M)$-POVMs in $d\geq 3$. In $d=2$, this number drops to two: general SIC POVMs and mutually unbiased measurements. Therefore, $(N,M)$-POVMs can be viewed as collections of mutually
unbiased symmetric POVMs. Partial conditions for the existence of symmetric measurements for the maximal value of $x$ have recently been proposed \cite{Alber3}.

In what follows, we consider a natural generalization of $(N,M)$-POVMs by taking a collection $\mathcal{E}=\{E_{\alpha,k};\, k=1,\ldots,M_\alpha;\,\alpha=1,\ldots,N\}$ of $N$  POVMs that no longer have to be equinumerous (in general, $M_\alpha\neq M_\beta$ for $\alpha\neq\beta$). Now, the symmetry conditions depend on the chosen POVM $\alpha$ but not on its element $k$.

\begin{Definition}
A generalized symmetric measurement is a set of $N$ POVMs $\mathcal{E}_\alpha=\{E_{\alpha,k};\, k=1,\ldots,M_\alpha\}$ such that
\begin{equation}\label{MM}
\begin{split}
\Tr (E_{\alpha,k})&=w_\alpha,\\
\Tr (E_{\alpha,k}^2)&=x_\alpha,\\
\Tr (E_{\alpha,k}E_{\alpha,\ell})&=y_\alpha,\qquad \ell\neq k,\\
\Tr (E_{\alpha,k}E_{\beta,\ell})&=z_{\alpha\beta},\qquad \beta\neq\alpha.
\end{split}
\end{equation}
\end{Definition}

In Appendix \ref{AppA}, we show that the defining parameters are constrained by
\begin{equation}
w_\alpha=\frac{d}{M_\alpha},\qquad y_\alpha=\frac{d-M_\alpha x_\alpha}{M_\alpha(M_\alpha-1)},\qquad z_{\alpha\beta}=\frac{d}{M_\alpha M_\beta},
\end{equation}
where the free parameters $x_\alpha$ belong to the range
\begin{equation}
\frac{d}{M_\alpha^2}<x_\alpha\leq \min\left\{\frac{d^2}{M_\alpha^2},\frac {d}{M_\alpha}\right\}.
\end{equation}
Additionally, the total number of measurement operators
\begin{equation}\label{num}
\sum_{\alpha=1}^NM_\alpha\leq d^2+N-1
\end{equation}
is constrained by the dimension $d$ and the number of POVMs $N$. The equality is reached if and only if the generalized symmetric measurement is informationally complete, which is proven in Appendix \ref{AppB}.

Note that the generalized symmetric measurement can be understood as a collection of $N$ $(1,M_\alpha)$-POVMs, each characterized by its own parameter $x_\alpha$. The $(N,M)$-POVMs are recovered if all $M_\alpha=M$ and $x_\alpha=x$. Therefore, there exists a subclass of equinumerous generalized symmetric measurements that are not $(N,M)$-POVMs due to $x_\alpha$ being dependent on the choice of a POVM $\mathcal{E}_\alpha$.

\begin{Example}\label{M12}
Contrary to $(N,M)$-POVMs, the generalized symmetric measurements introduce one new informationally complete class for $d=2$. It corresponds to the choice $M_1=2$ and $M_2=3$. An example of projective measurements consists in $E_{1,1}={\rm diag}(1,0)$, $E_{1,2}={\rm diag}(0,1)$, as well as
\begin{equation}
E_{2,1}=\frac 13 \begin{pmatrix} 1 & -i \\ i & 1 \end{pmatrix},\qquad
E_{2,2}=\frac 16 \begin{pmatrix} 2 & i+\sqrt{3} \\ -i+\sqrt{3} & 2  \end{pmatrix},\qquad
E_{2,3}=\frac 16 \begin{pmatrix} 2 & i-\sqrt{3} \\ -i-\sqrt{3} & 2
\end{pmatrix}.
\end{equation}
\end{Example}

\section{Construction methods}

To construct informationally complete generalized symmetric measurements, let us take a Hermitian orthonormal operator basis $\{G_0=\mathbb{I}_d/\sqrt{d},G_{\alpha,k};\,\alpha=1,\ldots,N,\,k=1,\ldots,M_\alpha-1\}$  with traceless $G_{\alpha,k}$. We use it to define another set of traceless operators,
\begin{equation}\label{H}
H_{\alpha,k}=\left\{\begin{aligned}
&G_\alpha-\sqrt{M_\alpha}(1+\sqrt{M_\alpha})G_{\alpha,k},\quad k=1,\ldots,M_\alpha-1,\\
&(1+\sqrt{M_\alpha})G_\alpha,\qquad k=M_\alpha,
\end{aligned}\right.
\end{equation}
where $G_\alpha=\sum_{k=1}^{M_\alpha-1}G_{\alpha,k}$, so that $\sum_{\alpha=1}^N\sum_{k=1}^{M_\alpha}H_{\alpha,k}=0$.
The measurement operators follow from
\begin{equation}\label{E}
E_{\alpha,k}=\frac{1}{M_\alpha} \mathbb{I}_d+t_\alpha H_{\alpha,k},
\end{equation}
where $t_\alpha$ is related to $x_\alpha$ via
\begin{equation}\label{t}
t_\alpha^2=\frac{M_\alpha^2x_\alpha-d}{M_\alpha^2(M_\alpha-1)(1+\sqrt{M_\alpha})^2}.
\end{equation}
Moreover, the positivity condition $E_{\alpha,k}\geq 0$ imposes the following constraint on $t_\alpha$,
\begin{equation}\label{trange}
-\frac{1}{M_\alpha}\frac{1}{\lambda_{\alpha,\max}}\leq t_\alpha\leq \frac{1}{M_\alpha}\frac{1}{|\lambda_{\alpha,\min}|},
\end{equation}
where $\lambda_{\alpha,\max}$ and $\lambda_{\alpha,\min}$ are the minimal and maximal eigenvalues from among all eigenvalues of $H_{\alpha,k}$, for a fixed $\alpha$.
It is straightforward to recover the inverse relation. That is, if the measurement operators are known, then the associated orthonormal Hermitian basis operators read
\begin{equation}
G_{\alpha,k}= \frac{1}{t_\alpha M_\alpha(1+\sqrt{M_\alpha})^2}\Big[\mathbb{I}_d
+\sqrt{M_\alpha}E_{\alpha,M_\alpha}-\sqrt{M}(1+\sqrt{M})E_{\alpha,k}\Big].
\end{equation}

\begin{Remark}
Actually, there is a non-empty range of $t_\alpha$, where $\pm t_\alpha$ correspond to the same value of $x_\alpha$. Following eq. (\ref{trange}), let us denote the boundary $t_\alpha$ by
\begin{equation}
t_{\alpha_\ast}=\frac{1}{M_\alpha}\frac{1}{\max\left\{\lambda_{\max},|\lambda_{\min}|\right\}}.
\end{equation}
In this case,
\begin{equation}\label{Epm}
E_{\alpha,k}^{(\pm)}=\frac{1}{M_\alpha} \mathbb{I}_d\pm t_{\alpha} H_{\alpha,k}
\end{equation}
are both valid generalized symmetric measurements for a positive $t_\alpha\leq t_{\alpha,\ast}$. Observe that
\begin{equation}
\begin{split}
\Tr E_{\alpha,k}^{(+)}E_{\alpha,k}^{(-)} & = 2z_{\alpha\alpha}-x_\alpha,\\
\Tr E_{\alpha,k}^{(+)}E_{\alpha,\ell}^{(-)} & = y_\alpha-\frac{2}{M_\alpha-1}z_{\alpha\alpha},\qquad k\neq\ell.
\end{split}
\end{equation}
In particular, $E_{\alpha,k}^{(+)}$ and $E_{\alpha,k}^{(-)}$ are orthogonal in the Hilbert-Schmidt inner product for $x_\alpha=2d/M_\alpha^2$.
\end{Remark}

\begin{Example}\label{mub}
As a simple example of mutually orthogonal $E_{\alpha,k}^{(+)}$ and $E_{\alpha,k}^{(-)}$, consider the POVM with $M_\alpha=2$ measurement operators, characterized by $x_\alpha=d/M_\alpha$. Then, $y_\alpha=0$ and
\begin{equation}
\begin{split}
\Tr E_{\alpha,k}^{(+)}E_{\alpha,k}^{(-)} & = 0,\\
\Tr E_{\alpha,k}^{(+)}E_{\alpha,\ell}^{(-)} & = \frac{d}{2}.
\end{split}
\end{equation}
In the special case of $d=2$ (MUBs), $\{E_{\alpha,k}^{(+)}\}=\{E_{\alpha,k}^{(-)}\}$.
\end{Example}

\begin{Example}\label{exx}
In $d=3$, let us take two diagonal elements of a Hermitian orthonormal basis \cite{Alber3},
\begin{equation}
G_1=\frac{1}{\sqrt{3}(\sqrt{3}+1)}
\begin{pmatrix}
-2-\sqrt{3} & 0 & 0 \\
0 & 1 & 0 \\
0 & 0 & 1+\sqrt{3}
\end{pmatrix},
\qquad
G_2=\frac{1}{\sqrt{3}(\sqrt{3}+1)}
\begin{pmatrix}
1 & 0 & 0 \\
0 & -2-\sqrt{3} & 0 \\
0 & 0 & 1+\sqrt{3}
\end{pmatrix},
\end{equation}
and construct two $(1,3)$-POVMs $\mathcal{E}_\pm=\{E_{k,\pm};\,k=1,2,3\}$ using eq. (\ref{E}) (the index $\alpha$ is dropped). For $t>0$, take $t=t_{\max}=\frac{\sqrt{3}-1}{2\sqrt{3}}$, so that the corresponding measurement operators
\begin{equation}\label{ep}
E_{1,+}={\rm diag}(1,0,0),\qquad E_{2,+}={\rm diag}(0,1,0),\qquad
E_{3,+}={\rm diag}(0,0,1)
\end{equation}
are projectors onto the operational basis ($x_+=1$). For $t<0$, take instead $t=t_{\min}=\frac{1-\sqrt{3}}{4\sqrt{3}}$, for which
\begin{equation}\label{em}
E_{1,-}=\frac 12 {\rm diag}(0,1,1),\qquad E_{2,-}=\frac 12 
{\rm diag}(1,0,1),\qquad E_{3,-}=\frac 12 {\rm diag}(1,1,0)
\end{equation}
are rank-2 projectors ($x_-=1/2$) and also orthonormal completions of respective $E_{k,+}$ to the identity operator. 
Note that $t_{\max}\geq|t_{\min}|$, and hence, even though we use the same formulas for $H_{\alpha,k}$, $t_{\max}$ allows for $x_+>x_-$. However, we can construct
\begin{equation}
\widetilde{E}_{k,\pm}=\frac{1}{3}\mathbb{I}_3\pm t_\ast H_{\alpha,k},
\qquad t_\ast=\min\{t_{\max},|t_{\min}|\}=|t_{\min}|.
\end{equation}
Obviously, $\widetilde{E}_{k,-}=E_{k,-}$ remain unchanged. However, we have found another set of rank-2 projectors
\begin{equation}
\widetilde{E}_{1,+}=\frac 16 {\rm diag}(4,1,1),\qquad
\widetilde{E}_{2,+}=\frac 16 {\rm diag}(1,4,1),\qquad
\widetilde{E}_{1,+}=\frac 16 {\rm diag}(1,1,4)
\end{equation}
characterized by $x_\ast=x_-=1/2$.
\end{Example}

The above method is in full analogy with the construction of $(N,M)$-POVMs \cite{SIC-MUB}, which is based on the construction methods for general SIC POVMs \cite{Gour} and mutually unbiased measurements \cite{Kalev}. In these works, the correspondence between the measurement operators $E_{\alpha,k}$ and the basis operators $G_{\alpha,k}$ is stated to be one-to-one. However, as already seen in Example \ref{mub}, this statement is incorrect. Moreover, there exists yet another correspondence based on an alternative way to introduce the operators $H_{\alpha,k}$ for the same set of $G_{\alpha,k}$. Namely, one defines
\begin{equation}\label{Hprime}
H_{\alpha,k}^\prime=\left\{\begin{aligned}
&G_\alpha+\sqrt{M_\alpha}(1-\sqrt{M_\alpha})G_{\alpha,k},\quad k=1,\ldots,M_\alpha-1,\\
&(1-\sqrt{M_\alpha})G_\alpha,\qquad k=M_\alpha.
\end{aligned}\right.
\end{equation}
We denote the associated measurement operators by
\begin{equation}\label{Eprime}
E_{\alpha,k}^\prime=\frac{1}{M_\alpha} \mathbb{I}_d+t_\alpha^\prime H_{\alpha,k}^\prime,
\end{equation}
where the relation between the parameter $t_\alpha^\prime$ and $x_\alpha$ reads
\begin{equation}\label{tprime}
t_\alpha^{\prime 2}
=\frac{M_\alpha^2x_\alpha-d}{M_\alpha^2(M_\alpha-1)(1-\sqrt{M_\alpha})^2}.
\end{equation}
Observe that $t_\alpha^\prime\neq t_\alpha$ but instead $t_\alpha^{\prime 2}(1-\sqrt{M_\alpha})^2=t_\alpha^2(1+\sqrt{M_\alpha})^2$.
For the sake of completion, we present the formulas for $G_{\alpha,k}$ in terms of $E_{\alpha,k}^\prime$;
\begin{equation}
G_{\alpha,k}= \frac{1}{t_\alpha^\prime M_\alpha(1-\sqrt{M_\alpha})^2}\Big[\mathbb{I}_d
-\sqrt{M_\alpha}E_{\alpha,M_\alpha}^\prime+\sqrt{M}(1-\sqrt{M})E_{\alpha,k}^\prime\Big].
\end{equation}

Now, let us compare the measurement $\mathcal{E}_\alpha$ and $\mathcal{E}_\alpha^\prime=\{E_{\alpha,k}^\prime;\, k=1,\ldots,M_\alpha\}$ that have been constructed using eqs. (\ref{E}) and (\ref{Eprime}), respectively.

\begin{Proposition}
Traceless orthonormal Hermitian operators $\{G_{\alpha,k};\,k=1,\ldots,M_\alpha\}$ produce two POVMs $\mathcal{E}_\alpha$, $\mathcal{E}_\alpha^\prime$ with the same elements if and only if $M_\alpha\leq 3$ and ${\rm sgn}(t_\alpha)=-{\rm sgn}(t_\alpha^\prime)$.
\end{Proposition}

The proof is given in Appendix \ref{AppC}. Example \ref{exx} provides a simple subset of orthonormal Hermitian operator basis that construct $\mathcal{E}_\alpha=\mathcal{E}_\alpha^\prime$ but with $E_{\alpha,k}\neq E_{\alpha,k}^\prime$.

An analogical reasoning to that of Remark 1 also follows for $E_{\alpha,k}^\prime$. This means that, for $M_\alpha\geq 4$ and sufficiently small $x_\alpha$, there exist a total of four ways of constructing $(1,M_\alpha)$-POVMs from the same $\{G_{\alpha,k};\,k=1,\ldots,M_\alpha\}$. Moreover, each $(1,M_\alpha)$-POVM inside the same generalized symmetric measurement can be constructed using a different method due to $H_{\alpha,k}$ and $H_{\beta,\ell}^\prime$ being mutually orthogonal for $\beta\neq\alpha$.

\begin{Remark}
The presented construction methods provide up to $4N$ distinct (generalized) symmetric measurements that arise from the same partition $\{G_{\alpha,k};\,k=1,\ldots,M_\alpha;\,\alpha=1,\ldots,N\}$ of traceless orthonormal Hermitian operators. 
\end{Remark}

An example for $N=1$ is presented in Appendix \ref{AppD}. Whether there exist unequivalent constructions that use a parametrization of $H_{\alpha,k}$ different from
\begin{equation}
H_{\alpha,k}=\left\{\begin{aligned}
&G_\alpha+AG_{\alpha,k},\quad k=1,\ldots,M_\alpha-1,\\
&BG_\alpha,\qquad k=M_\alpha,
\end{aligned}\right.
\end{equation}
where $A$ and $B$ are constants, is an open question.

\section{Recovering symmetries of measurements}

Due to their high symmetry, the $(N,M)$-POVMs are characterized via only four parameters: $(w,x,y,z)$. In the process of generalization to the generalized symmetric measurements, which are collections of POVMs that are in general unequinumerous, some of this symmetry is lost. In particular, one now needs $N$ sets of parameters $(w_\alpha,x_\alpha,y_\alpha,z_{\alpha\beta})$ to characterize them. Luckily, there are certain cases in which it is possible to recover some of the symmetry conditions of the $(N,M)$-POVMs. In Section 2, we have already discussed equinumerous measurements ($M_\alpha=M$), which correspond to $w_\alpha=w$ and $z_{\alpha,\beta}=z$. On the contrary, if $z_{\alpha\beta}=z$, then equinumerous measurements follow only for $N\geq 3$, which will allow for further symmetrization of other special classes that are listed below.

\begin{itemize}
\item If $\max\{d/M_\alpha^2\}<x\leq\min\{d/M_\alpha,d^2/M_\alpha^2\}$ is a non-empty set, then there exist generalized symmetric measurements for which $x_\alpha=x$. However, $y_\alpha=y_\beta$ only for $M_\alpha=M_\beta$.
\item If $\max\{0,\frac{d(M_\alpha-d)}{M_\alpha^2(M_\alpha-1)}\}<y\leq\min\{d/M_\alpha^2\}$ is a non-empty set, then there exist generalized symmetric measurements such that $y_\alpha=y$. In this case, $x_\alpha=w_\alpha-y(M_\alpha-1)\neq x_\beta$ for $M_\alpha\neq M_\beta$.
\end{itemize}

\begin{Example}\label{M23}
In $d=2$, the only unequinumerous generalized symmetric measurements that are also informationally complete follow from $M_1=2$ and $M_2=3$. Note that $z_{12}=z_{21}=1/3$ is constant, even though $w_1\neq w_2$. As unequinumerous measurements allow for at most two $\alpha$-independent parameters, we now consider two distinct classes.
\begin{enumerate}[label=(\roman*)]
\item There is no measurement with $x_1=x_2=x$ due to $1/2<x\leq 4/9$ being an empty set. However, already for $d=3$, the range $3/4<x\leq 1$ of $x$ is non-empty.
\item There exist measurements with $y_1=y_2=y$, where $1/9<y\leq 2/9$. In this case, $x_1=1-y$ and $x_2=2/3-2y$.
\end{enumerate}
\end{Example}

In the upcoming section, it will become evident that other forms of symmetry are even more important for the applicational purposes.

\begin{Proposition}\label{Er}
If the generalized symmetric measurement is characterized by
\begin{equation}
x_\alpha=\frac{d+rM_\alpha(M_\alpha-1)}{M_\alpha^2},\qquad 
y_\alpha=\frac{d-rM_\alpha}{M_\alpha^2}
\end{equation}
with the constant parameter $r$ from the range
\begin{equation}
0<r\leq\min\left\{\frac{d}{M_\alpha},\frac{d(d-1)}{M_\alpha(M_\alpha-1)}\right\},
\end{equation}
then $x_\alpha-y_\alpha=r$.
\end{Proposition}

\begin{Proposition}\label{Es}
If the generalized symmetric measurement is characterized by
\begin{equation}
x_\alpha=\frac{d}{M_\alpha^2}[1+s(M_\alpha-1)],\qquad y_\alpha=\frac{d}{M_\alpha^2}(1-s),
\end{equation}
where the parameter $s$ satisfies
\begin{equation}
0<s\leq\min\left\{1,\frac{d-1}{M_\alpha-1}\right\},
\end{equation}
then $x_\alpha-y_\alpha=sw_\alpha$.
\end{Proposition}

In particular, for the class of generalized symmetric measurements from Example \ref{M23}, the maximal value of $x_\alpha-y_\alpha=r$ is $r_{\max}=1/3$. On the other hand, if $x_\alpha-y_\alpha=sw_\alpha$, then $s_{\max}=1/2$.

\section{Applications}

\subsection{Conical 2-designs}

Complex projective 2-designs are families or rank-1 projectors $P_k$, where $\sum_kP_k\otimes P_k$ commutes with $U\otimes U$ for any unitary operator $U$. Important examples include SIC POVMs and mutually unbiased bases. Their applications include quantum state tomography \cite{Adamson,Scott}, quantum key distribution \cite{Renes2,Cr1}, and quantum entanglement detection \cite{Spengler,ESIC}. Appleby and Graydon introduced a generalization to conical 2-designs by replacing rank-1 projectors $P_k$ with positive operators $E_k$ \cite{Graydon,Graydon2}. From definition, it follows that $E_k$ are conical 2-designs if and only if
\begin{equation}\label{con}
\sum_kE_k\otimes E_k=
\kappa_+\mathbb{I}_d\otimes\mathbb{I}_d+\kappa_-\mathbb{F}_d,
\end{equation}
where $\kappa_+\geq\kappa_->0$ and $\mathbb{F}_d=\sum_{m,n=1}^d|m\>\<n|\otimes|n\>\<m|$ is the flip operator \cite{Graydon}. Known conical 2-designs include general SIC POVMs and mutually unbiased measurements \cite{Wang}. Recently, it has been shown that informationally complete $(N,M)$-POVMs are also conical 2-designs \cite{SICMUB_design,SICMUB_channels}.

Let us show that eq. (\ref{con}) is satisfied for even more general measurements. We start by introducing $N$ linear maps
\begin{equation}
\Phi_\alpha[X]=\sum_{k=1}^{M_\alpha}E_{\alpha,k}\Tr(XE_{\alpha,k}),
\end{equation}
the maximally depolarizing channel $\Phi_0[X]=\mathbb{I}_d\Tr(X)/d$, and the identity operator $\oper$. In Appendix \ref{AppE}, we prove that if
\begin{equation}\label{suma}
\sum_{\alpha=1}^N\Phi_\alpha=\kappa_-\oper+\kappa_+d\Phi_0,
\end{equation}
then the corresponding $E_{\alpha,k}$ are a conical 2-design
\begin{equation}
\sum_{\alpha=1}^N\sum_{k=1}^{M_\alpha}E_{\alpha,k}\otimes E_{\alpha,k}
=\kappa_+\mathbb{I}_d\otimes\mathbb{I}_d+\kappa_-\mathbb{F}_d
\end{equation}
characterized by the same values of $\kappa_\pm$. However, for arbitrary generalized symmetric measurements, the condition in eq. (\ref{suma}) turns out to be too strong. It becomes necessary to impose the additional constraints on the measurement operators.

\begin{Proposition}\label{c2d}
If $E_{\alpha,k}$ belong to the class of measurements from Proposition \ref{Er}, for which $x_\alpha-y_\alpha=r$, then they are a conical 2-design with
\begin{equation}\label{kappas2}
\kappa_+=\mu-\frac rd,\qquad\kappa_-=r.
\end{equation}
\end{Proposition}

For the proof, see Appendix \ref{AppF}. Analogical calculations can be repeated for the measurements from Proposition \ref{Es}, for which instead $x_\alpha-y_\alpha=sw_\alpha$. In this case, one first constructs entanglement breaking channels $\widetilde{\Phi}_\alpha$ such that
\begin{equation}
\widetilde{\Phi}_\alpha[X]=\sum_{k=1}^{M_\alpha}\frac{1}{w_\alpha}E_{\alpha,k}\Tr(E_{\alpha,k}X),\qquad
\sum_{\alpha=1}^N\widetilde{\Phi}_\alpha=s\oper+(N-s)\Phi_0.
\end{equation}
However, as it turns out, these measurements are not conical 2-designs. Rather than eq. (\ref{con}), they satisfy
\begin{equation}
\sum_{\alpha=1}^N\sum_{k=1}^{M_\alpha}\frac{1}{w_\alpha}E_{\alpha,k}\otimes E_{\alpha,k}
=\kappa_+\mathbb{I}_d\otimes\mathbb{I}_d + \kappa_- \mathbb{F}_d,
\end{equation}
which reduces to conical 2-designs only in a very special case of mutually unbiased bases ($w_\alpha=1$). For more details, see Appendix \ref{AppG}.

\subsection{Index of coincidence}

From definition, the index of coincidence is a sum of squared probabilities \cite{Rastegin5}. For the generalized symmetric measurements, one has
\begin{equation}
C=\sum_{\alpha=1}^N\sum_{k=1}^{M_\alpha}p_{\alpha,k}^2,
\end{equation}
where the probability distributions $\{p_{\alpha,k};\,k=1,\ldots,M_\alpha\}$ follow from the expansion of a mixed state
\begin{equation}
\rho=\sum_{\alpha=1}^N\sum_{k=1}^{M_\alpha}p_{\alpha,k}F_{\alpha,k},
\end{equation}
in the dual frame (see Appendix \ref{AppH})
\begin{equation}\label{Fak}
F_{\alpha,k}=\frac{1}{x_\alpha-y_\alpha}\left[E_{\alpha,k}-\frac 1d \mathbb{I}_d\left(w_\alpha-\frac{x_\alpha-y_\alpha}{N}\right)\right].
\end{equation}
Calculating the index of coincidence is crucial in order to derive entropic uncertainty relations and separability criteria, as will be seen in the upcoming subsections. Once again, it turns out that obtaining an analytical formula for $C$ is possible only for a special class of measurements.

\begin{Proposition}\label{prop}
If $E_{\alpha,k}$ belong to the class of measurements from Proposition \ref{Er}, for which $x_\alpha-y_\alpha=r$, then the index of coincidence $C$ is bounded by
\begin{equation}
C\leq C_{\max}=\frac{d-1}{d}r+\mu,
\end{equation}
where $\mu=\sum_{\alpha=1}^NM_\alpha^{-1}$.
\end{Proposition}

The proof is provided in Appendix \ref{AppI}. The upper bound $C_{\max}$ is reached on pure states.

\subsection{Entropic uncertainty relations}

Uncertainty relations are one of the fundamental concepts in quantum theory. In quantum information, they are ofter expressed in terms of entropies for probability distributions associated with measurement operators. Entropic uncertainty relations have important applications e.g. in quantum cryptography \cite{Koashi,Coles} and quantum entanglement \cite{Guhne2,Rastegin4}. An important problem is finding bounds for sums of entropies \cite{AEUR,MEUR}.

We derive entropic uncertainty relations for the generalized symmetric measurements in terms of the index of coincidence. For the Shannon entropy $H(\mathcal{E}_\alpha,\rho)=-\sum_{k=1}^{M_\alpha}p_{\alpha,k}\log p_{\alpha,k}$ associated with the measurement $\mathcal{E}_\alpha=\{E_{\alpha,k};\,k=1,\ldots,M_\alpha\}$, one finds
\begin{equation}\label{ent}
\frac{1}{N}\sum_{\alpha=1}^NH(\mathcal{E}_\alpha,\rho)
\geq\log\frac{N}{C}.
\end{equation}
The proof is analogical to this in ref. \cite{SIC-MUB}. It uses the concavity of logarithms and Jensen's inequality $H(\mathcal{E}_\alpha,\rho)\geq R(\mathcal{E}_\alpha,\rho)$ \cite{Sanchez,Maassen}, where the R\'{e}nyi 2-entropy $R(\mathcal{E}_\alpha,\rho)=-\log \sum_{k=1}^{M_\alpha}p_{\alpha,k}^2$ \cite{Renyi}.

In particular, if $d=2$, there is the following relation for the lower bounds from eq. (\ref{ent}),
\begin{equation}
\left.\frac{N}{C_{\max}}\right|_{\rm SIC}=\log 3\geq
\left.\frac{N}{C_{\max}}\right|_{\rm GSM}=\log 2\geq
\left.\frac{N}{C_{\max}}\right|_{\rm MUB}=\log \frac 32,
\end{equation}
where GSM are the generalized symmetric measurements from Example \ref{M23} with the maximal value of $r=1/3$. Interestingly, a pair of unequinumerous measurements corresponds to a higher entropic bound than a triple of mutually unbiased bases.

\subsection{Separability criteria}

Quantum entanglement is a type of non-classical correlations and a crucial feature in quantum theory. It becomes a useful resource for many quantum tasks, e.g. quantum computation \cite{Raus}, quantum communication \cite{Piveteau}, quantum cryptography \cite{Ekert}, or quantum teleportation \cite{Brassard}. For this reason, it is important to develop methods to detect entangled states and quantify the amount of entanglement.

Consider a bipartite state $\rho$ on the composite Hilbert space $\mathcal{H}=\mathcal{H}_A\otimes\mathcal{H}_B$, where $\dim\mathcal{H}_{A/B}=d_{A/B}$. For each subsystem, let us introduce generalized symmetric measurements and denote them by $\{E_{\alpha,k}^{A/B}\}$. Now, linear correlations between $\{E_{\alpha,k}^{A}\}$ and $\{E_{\alpha,k}^{B}\}$ are encoded in the correlation matrix
\begin{equation}
\mathcal{P}_{\alpha,k;\beta,\ell}=\Tr\Big[\rho(E_{\alpha,k}^{A}\otimes E_{\beta,\ell}^{B})\Big].
\end{equation}
This matrix can be used to formulate two necessary conditions for separability of $\rho$.

\begin{Proposition}\label{esic}
If a bipartite state $\rho$ is separable, then
\begin{align}
\Tr\mathcal{P}\leq\frac{C_{\max}^{A}+C_{\max}^{B}}{2}&\qquad{\rm for}\quad d_A=d_B,\label{trP}\\
\|\mathcal{P}\|_{\Tr}\leq \sqrt{C_{\max}^{A}C_{\max}^{B}}&\qquad{\rm for\,any}\quad d_A,d_B,\label{normP},
\end{align}
where $C_{\max}^{A/B}$ is the upper bound for the index of coincidence corresponding to $\{E_{\alpha,k}^{A/B}\}$.
\end{Proposition}

This is proven exactly as in ref. \cite{SIC-MUB}: by computing $\Tr\mathcal{P}$ and $\|\mathcal{P}\|_{\Tr}$ on product states and then extending the results to all separable states (from the convexity of the trace norm and linearity of trace). For $d_A=d_B$, it remains inconclusive which of the two conditions is stronger. In the special case where $C_{\max}^{A}=C_{\max}^{A}$, both upper bounds coincide. Also, the smaller the values of $r=x_\alpha-y_\alpha$ and $\mu=\sum_{\alpha=1}^NM_\alpha^{-1}$, the tighter these bounds are. This means that taking nonprojective, unequinumerous measurements proves to be more beneficial for entanglement detection.



\section{Conclusions}

In this paper, we introduce generalized symmetric measurements, which are collections of $M_\alpha$-elemental POVMs $\mathcal{E}_\alpha$ that satisfy additional symmetry constraints. In general these POVMs are unequinumerous, and hence mutual unbiasedness is preserved only between pairs. For informationally complete measurements, we present construction methods from Hermitian orthonormal operator bases. In particular, we find that a single $M_\alpha$-elemental subset of Hermitian orthonormal bases can produce up to four distinct POVMs $\mathcal{E}_\alpha$, which was believed to be a one-to-one correspondence \cite{Kalev,Gour,SIC-MUB}. In the main part, we prove that some of the symmetry conditions of $(N,M)$-POVMs can be recovered for their generalizations. Interestingly, the class with the greatest potential for applications is characterized not by a single measurement-independent parameter but by a difference;
\begin{equation}
x_\alpha-y_\alpha=\Tr\Big[E_{\alpha,k}(E_{\alpha,k}-E_{\alpha,\ell})\Big]\equiv r\qquad \bigforall_{k\neq\ell}.
\end{equation}
If the generalized symmetric measurements satisfy this condition, then they are conical 2-designs, and the corresponding index of coincidence is analytically computable. Finally, we present possible applications of such measurements in entropic uncertainty relations and separability criteria for bipartite quantum states.

In further research, it would be essential to further characterize the properties of generalized symmetric measurements. In particular, there might exist non-trivial relations between the POVMs constructed from the same Hermitian operator bases. It would be interesting to characterize the families of projective measurements and compare their applicational prowess with MUBs and SIC POVMs. Also, there is an open question about possible applications for the measurements that are not conical 2-designs.

\section{Acknowledgements}

This research was funded in whole or in part by the National Science Centre, Poland, Grant number 2021/43/D/ST2/00102. For the purpose of Open Access, the author has applied a CC-BY public copyright licence to any Author Accepted Manuscript (AAM) version arising from this submission.

\appendix

\section{Characterizing parameters}\label{AppA}

To obtain the values of parameters $w_\alpha$, $y_\alpha$, and $z_{\alpha\beta}$ that characterize the generalized symmetric measurements, we compute the trace conditions using decompositions of the identity $\mathbb{I}_d=\sum_{k=1}^{M_\alpha}E_{\alpha,k}$. For $w_\alpha$, one has
\begin{equation}
d=\Tr\mathbb{I}_d=\sum_{k=1}^{M_\alpha}\Tr E_{\alpha,k}=M_\alpha w_\alpha.
\end{equation}
Then, $z_{\alpha\beta}$ follows from
\begin{equation}
w_\alpha=\Tr E_{\alpha,k}=\sum_{\ell=1}^{M_\beta}\Tr E_{\alpha,k}E_{\beta,\ell}=M_\beta z_{\alpha\beta}.
\end{equation}
Finally, we obtain the relation between $x_\alpha$ and $y_\alpha$ from
\begin{equation}
w_\alpha=\Tr E_{\alpha,k}=\sum_{\ell=1}^{M_\alpha}\Tr E_{\alpha,k}E_{\alpha,\ell}=x_\alpha+(M_\alpha-1)y_\alpha.
\end{equation}
The range of $x_\alpha$ depends on the dimension $d$ and the number of elements $M_\alpha$ of the given POVM. The lower bound $x_\alpha=d/M_\alpha^2$ corresponds to the trivial choice of $E_{\alpha,k}=\mathbb{I}_d/M$. The upper bound $x_\alpha=d^2/M_\alpha^2$ is reached by projective measurements, whereas $x_\alpha=d/M_\alpha$ for rank-$w_\alpha$ projectors.

\section{Informationally complete measurements}\label{AppB}

Any set of measurements is informationally complete if and only if it consists in $d^2$ linearly independent operators. Because of the constraint $\sum_{k=1}^{M_\alpha}E_{\alpha,k}=\mathbb{I}_d$, every POVM $\{E_{\alpha,k};\,k=1,\ldots,M_\alpha\}$ contains at most $M_\alpha-1$ linearly independent operators. Therefore, the generalized symmetric measurements are spanned by the identity operator $\mathbb{I}_d$ and $d^2-1$ operators $E_{\alpha,k}$. This imposes the following condition on $M_\alpha$,
\begin{equation}
\sum_{\alpha=1}^N(M_\alpha-1)=d^2-1,
\end{equation}
which is exactly the equality in eq. (\ref{num}). Now, to show that $\{\mathbb{I}_d,E_{\alpha,k};\,k=1,\ldots,M_\alpha-1;\,\alpha=1,\ldots,N\}$ is indeed a set of linearly independent operators, it is enough to prove that
\begin{equation}\label{cond}
\mathbb{E}=r_0\mathbb{I}_d+\sum_{\alpha=1}^N\sum_{k=1}^{M_\alpha-1}r_{\alpha,k}E_{\alpha,k}=0
\end{equation}
implies $r_0=r_{\alpha,k}=0$. First, observe that
\begin{equation}\label{cond2}
\Tr(\mathbb{E})=dr_0+\sum_{\alpha=1}^N\sum_{k=1}^{M_\alpha-1}r_{\alpha,k}w_\alpha=0.
\end{equation}
Using this result, one obtains
\begin{equation}
\Tr(\mathbb{E}E_{\beta,M})=\sum_{k=1}^{M_\beta-1}r_{\beta,k}(y_\beta-z_{\beta\beta})=0.
\end{equation}
As $y_\beta\neq z_{\beta\beta}$ due to $x_\beta>d/M_\beta^2$, the only solution is
\begin{equation}\label{cond3}
\sum_{k=1}^{M_\beta-1}r_{\beta,k}=0.
\end{equation}
Now, eq. (\ref{cond3}) together with eq. (\ref{cond2}) return $r_0=0$. Finally, for any $\ell=1,\ldots,M_\beta-1$,
\begin{equation}
\Tr(\mathbb{E}E_{\beta,\ell})=r_{\beta,\ell}(x_\beta-y_\beta)=0,
\end{equation}
where $x_\beta\neq y_\beta$. Therefore, one indeed has $r_{\beta,\ell}=r_0=0$.

\section{Conditions for $\mathcal{E}_\alpha^\prime=\mathcal{E}_\alpha$}\label{AppC}

To prove that $E_{\alpha,k}$ and $E_{\alpha,k}^\prime$ constructed according to eqs. (\ref{E}) and (\ref{Eprime}), respectively, can indeed produce distinct measurement operators, it is enough to calculate the trace cross-relations
\begin{equation}
\Tr E_{\alpha,k}E_{\alpha,\ell}^\prime=\frac{d}{M_\alpha^2}+t_\alpha t_\alpha^\prime \Tr H_{\alpha,k}H_{\alpha,\ell}^\prime
=z_{\alpha\alpha}+{\rm sgn}(t_\alpha){\rm sgn}(t_\alpha^\prime)
\frac{x_\alpha-y_\alpha}{M_\alpha(M_\alpha-1)}\Tr H_{\alpha,k}H_{\alpha,\ell}^\prime,\qquad k,\ell=1,\ldots,M_\alpha,
\end{equation}
where
\begin{equation}\label{HH}
\begin{split}
\Tr H_{\alpha,k}H_{\alpha,k}^\prime &= M_\alpha^2-2M_\alpha-1,\\
\Tr H_{\alpha,M_\alpha}H_{\alpha,M_\alpha}^\prime &= -(M_\alpha-1)^2,\\
\Tr H_{\alpha,k}H_{\alpha,\ell}^\prime &= -(M_\alpha+1),\\
\Tr H_{\alpha,k}H_{\alpha,M_\alpha}^\prime &=
\Tr H_{\alpha,M_\alpha}H_{\alpha,k}^\prime = M_\alpha-1
\end{split}
\end{equation}
for $k,\ell=1,\ldots,M_\alpha-1$ and $k\neq\ell$. Because of the symmetry properties for the generalized symmetric measurements, some of the above traces have to be equal to one another. This is possible only for $M_\alpha\leq 3$.

First, let us see what happens if $M_\alpha=2$. Observe that the trace relations on eq. (\ref{HH}) reduce to
\begin{equation}
\begin{split}
\Tr H_{\alpha,1}H_{\alpha,1}^\prime &= \Tr H_{\alpha,2}H_{\alpha,2}^\prime=-1,\\
\Tr H_{\alpha,1}H_{\alpha,2}^\prime &=\Tr H_{\alpha,2}H_{\alpha,1}^\prime = 1.
\end{split}
\end{equation}
Therefore, the relations between measurement operators
\begin{equation}
\begin{split}
\Tr E_{\alpha,1}E_{\alpha,1}^\prime
&=\Tr E_{\alpha,2}E_{\alpha,2}^\prime
=\frac{d}{4}-{\rm sgn}(t_\alpha){\rm sgn}(t_\alpha^\prime)
\frac{4x_\alpha-d}{4}=x_\alpha,\\
\Tr E_{\alpha,1}E_{\alpha,2}^\prime
&=\Tr E_{\alpha,2}E_{\alpha,1}^\prime
=\frac{d}{4}+{\rm sgn}(t_\alpha){\rm sgn}(t_\alpha^\prime)
\frac{4x_\alpha-d}{4}=\frac{d-2x_\alpha}{2}=y_\alpha
\end{split}
\end{equation}
recover the symmetry conditions of the generalized symmetric measurements from eq. (\ref{MM}) when ${\rm sgn}(t_\alpha)=-{\rm sgn}(t_\alpha^\prime)$. Actually, one has $E_{\alpha,1}=E_{\alpha,1}^\prime$ and $E_{\alpha,2}=E_{\alpha,2}^\prime$.

Next, we fix the number of elements to $M_\alpha=3$. In this case, eq. (\ref{HH}) returns
\begin{equation}
\begin{split}
\Tr H_{\alpha,1}H_{\alpha,1}^\prime &=\Tr H_{\alpha,2}H_{\alpha,2}^\prime
=2,\\
\Tr H_{\alpha,3}H_{\alpha,3}^\prime &= -4,\\
\Tr H_{\alpha,1}H_{\alpha,2}^\prime &=\Tr H_{\alpha,2}H_{\alpha,1}^\prime 
=-4,\\
\Tr H_{\alpha,1}H_{\alpha,3}^\prime &=\Tr H_{\alpha,2}H_{\alpha,3}^\prime
=\Tr H_{\alpha,3}H_{\alpha,1}^\prime=\Tr H_{\alpha,3}H_{\alpha,2}^\prime
=2.
\end{split}
\end{equation}
For the measurement operators, it follows that
\begin{equation}
\begin{split}
\Tr E_{\alpha,1}E_{\alpha,2}^\prime
&=\Tr E_{\alpha,2}E_{\alpha,1}^\prime
=\Tr E_{\alpha,3}E_{\alpha,3}^\prime
=\frac{d}{9}-{\rm sgn}(t_\alpha){\rm sgn}(t_\alpha^\prime)
\frac{9x_\alpha-d}{9}=x_\alpha,\\
\Tr E_{\alpha,1}E_{\alpha,1}^\prime
&=\Tr E_{\alpha,2}E_{\alpha,2}^\prime
=\Tr E_{\alpha,1}E_{\alpha,3}^\prime =\Tr E_{\alpha,2}E_{\alpha,3}^\prime
=\Tr E_{\alpha,3}E_{\alpha,1}^\prime=\Tr E_{\alpha,3}E_{\alpha,2}^\prime\\
&=\frac{d}{9}+{\rm sgn}(t_\alpha){\rm sgn}(t_\alpha^\prime)
\frac{9x_\alpha-d}{18}=\frac{d-3x}{6}=y_\alpha,
\end{split}
\end{equation}
and hence the symmetry conditions from eq. (\ref{MM}) once again hold for ${\rm sgn}(t_\alpha)=-{\rm sgn}(t_\alpha^\prime)$. However, contrary to the case for $M_\alpha=2$, we have $E_{\alpha,1}=E_{\alpha,2}^\prime$, $E_{\alpha,2}=E_{\alpha,1}^\prime$, and $E_{\alpha,3}=E_{\alpha,3}^\prime$.

\section{Construction of four SIC POVMs in $d=2$}\label{AppD}

For qubit systems, SIC POVMs are recovered if $N=1$ and $M=d^2=4$. From the Pauli matrices $G_k=\sigma_k/\sqrt{2}$, eqs. (\ref{E}) and (\ref{Eprime}) lead to four families of projective measurements:
\begin{equation}
\left\{\begin{split}
E_{1,+}&=\frac{\sqrt{3}}{36}\begin{pmatrix} 1+3\sqrt{3} & -5-i \\
-5+i & -1+3\sqrt{3}
\end{pmatrix},\qquad
E_{2,+}=\frac{\sqrt{3}}{36}\begin{pmatrix} 1+3\sqrt{3} & 1+5i \\
1-5i & -1+3\sqrt{3}
\end{pmatrix},\\
E_{3,+}&=\frac{\sqrt{3}}{36}\begin{pmatrix} -5+3\sqrt{3} & 1-i \\
1+i & 5+3\sqrt{3}
\end{pmatrix},\qquad
E_{4,+}=\frac{\sqrt{3}}{12}\begin{pmatrix} 1+\sqrt{3} & 1-i \\
1+i & -1+\sqrt{3}
\end{pmatrix},
\end{split}\right.
\end{equation}
\begin{equation}
\left\{\begin{split}
E_{1,-}&=\frac{\sqrt{3}}{36}\begin{pmatrix} -1+3\sqrt{3} & 5+i \\
5-i & 1+3\sqrt{3}
\end{pmatrix},\qquad
E_{2,-}=\frac{\sqrt{3}}{36}\begin{pmatrix} -1+3\sqrt{3} & 5+i \\
5-i & 1+3\sqrt{3}
\end{pmatrix},\\
E_{3,-}&=\frac{\sqrt{3}}{36}\begin{pmatrix} 5+3\sqrt{3} & -1+i \\
-1-i & -5+3\sqrt{3}
\end{pmatrix},\qquad
E_{4,-}=\frac{\sqrt{3}}{12}\begin{pmatrix} -1+\sqrt{3} & -1+i \\
-1-i & 1+\sqrt{3}
\end{pmatrix},
\end{split}\right.
\end{equation}
\begin{equation}
\left\{\begin{split}
E_{1,+}^\prime&=\frac{\sqrt{3}}{12}\begin{pmatrix} 1+\sqrt{3} & -1-i \\
-1+i & -1+\sqrt{3}
\end{pmatrix},\qquad
E_{2,+}^\prime=\frac{\sqrt{3}}{12}\begin{pmatrix} 1+\sqrt{3} & 1+i \\
1-i & -1+\sqrt{3}
\end{pmatrix},\\
E_{3,+}^\prime&=\frac{\sqrt{3}}{12}\begin{pmatrix} -1+\sqrt{3} & 1-i \\
1+i & 1+\sqrt{3}
\end{pmatrix},\qquad
E_{4,+}^\prime=\frac{\sqrt{3}}{12}\begin{pmatrix} -1+\sqrt{3} & -1+i \\
-1-i & 1+\sqrt{3}
\end{pmatrix},
\end{split}\right.
\end{equation}
\begin{equation}
\left\{\begin{split}
E_{1,-}^\prime&=\frac{\sqrt{3}}{12}\begin{pmatrix} -1+\sqrt{3} & 1+i \\
1-i & 1+\sqrt{3}
\end{pmatrix},\qquad
E_{2,-}^\prime=\frac{\sqrt{3}}{12}\begin{pmatrix} -1+\sqrt{3} & -1-i \\
-1+i & 1+\sqrt{3}
\end{pmatrix},\\
E_{3,-}^\prime&=\frac{\sqrt{3}}{12}\begin{pmatrix} 1+\sqrt{3} & -1+i \\
-1-i & -1+\sqrt{3}
\end{pmatrix},\qquad
E_{4,-}^\prime=\frac{\sqrt{3}}{12}\begin{pmatrix} 1+\sqrt{3} & 1-i \\
1+i & -1+\sqrt{3}
\end{pmatrix},
\end{split}\right.
\end{equation}
where $E_{k,\pm}=E_k(\pm t_\ast)$ with $1/t_\ast=6\sqrt{6}$ and $1/t_\ast^\prime=2\sqrt{6}$. By reverse engineering the operator bases from these measurement operators, we find that any of these SIC POVMs can be constructed from $G_k=\pm\sigma_k/\sqrt{2}$ or $G_k=\pm g_k$ with
\begin{equation}
g_1=\frac{1}{3\sqrt{2}}\begin{pmatrix} 2 & -1-2i \\
-1+2i & -2
\end{pmatrix},\qquad
g_2=\frac{1}{3\sqrt{2}}\begin{pmatrix} 2 & 2+i \\
2-i & -2
\end{pmatrix},\qquad
g_3=\frac{1}{3\sqrt{2}}\begin{pmatrix} -1 & 2-2i \\
2+2i & 1
\end{pmatrix}.
\end{equation}

\section{Conical 2-designs}\label{AppE}

From the Choi-Jamio{\l}kowski isomorphism, we construct the Choi map
\begin{equation}
C(\Phi T)=\sum_{\alpha=1}^N(\oper_d\otimes \Phi_\alpha T)[dP_+]
\end{equation}
for $\Phi=\sum_{\alpha=1}^N\Phi_\alpha$, where $P_+=(1/d)\sum_{m,n=1}^d|m\>\<n|\otimes|m\>\<n|$ is the maximally entangled state. Direct calculations show that
\begin{equation}
C(\Phi T)=
\sum_{\alpha=1}^N\sum_{m,n=1}^d|m\>\<n|\otimes\Phi_\alpha[|n\>\<m|]
=\sum_{\alpha=1}^N\sum_{k=1}^{M_\alpha}\sum_{m,n=1}^d
|m\>\<m|E_{\alpha,k}|n\>\<n|\otimes E_{\alpha,k}=
\sum_{\alpha=1}^N\sum_{k=1}^{M_\alpha}E_{\alpha,k}\otimes E_{\alpha,k}.
\end{equation}
On the other hand, using the relation from eq. (\ref{suma}), one finds
\begin{equation}
C(\Phi T)=\left(\oper_d\otimes\Big[\kappa_+d\Phi_0+\kappa_-T\Big]\right)[dP_+]
=\sum_{m,n=1}^d|m\>\<n|\otimes\Big(\kappa_+\mathbb{I}_d\delta_{mn}+\kappa_-|n\>\<m|\Big)
=\kappa_+\mathbb{I}_d\otimes\mathbb{I}_d+\kappa_-\mathbb{F}_d.
\end{equation}
By comparing the above results, we finally obtain
\begin{equation}
\sum_{\alpha=1}^N\sum_{k=1}^{M_\alpha}E_{\alpha,k}\otimes E_{\alpha,k}
=\kappa_+\mathbb{I}_d\otimes\mathbb{I}_d + \kappa_- \mathbb{F}_d,
\end{equation}
and therefore $E_{\alpha,k}$ is a conical 2-design.

\section{Derivation of $\kappa_\pm$}\label{AppF}

Multiplying $\Phi[X]=\sum_{\alpha=1}^N\Phi_\alpha[X]$ by $E_{\beta,\ell}$ and taking the trace results in
\begin{equation}
\Tr(\Phi[X]E_{\beta,\ell})=\sum_{\alpha=1}^N\Tr(\Phi_\alpha[X]E_{\beta,\ell})
=(x_\beta-y_\beta)\Tr(E_{\beta,\ell}X)
+(y_\beta-z_{\beta\beta}+\mu w_\beta)\Tr X.
\end{equation}
Equivalently, this can be rewritten into the following relation,
\begin{equation}\label{trace}
\Tr\left\{\left(\Phi-\Big[(x_\beta-y_\beta)\oper
+(d\mu-x_\beta+y_\beta)\Phi_0\Big]\right)[X]E_{\beta,\ell}\right\}=0,
\end{equation}
which must hold for any Hermitian $X$ and $E_{\beta,\ell}$. It is easy to notice that the terms multiplying $\Tr X$ and $\Tr E_{\beta,\ell}X$ have to be independent on $\beta$ or $\ell$. This is true only for $x_\alpha-y_\alpha=r$. After substituting this into eq. (\ref{trace}), it follows that
\begin{equation}
\sum_{\alpha=1}^N\Phi_\alpha=r\oper+(d\mu-r)\Phi_0
=\kappa_-\oper+\kappa_+d\Phi_0,
\end{equation}
where $\kappa_{\pm}$ are given by eq. (\ref{kappas2}).

From Proposition \ref{Er}, we recall the range of the parameter $r$,
\begin{equation}
0<r\leq\min\left\{\frac{d}{M_\alpha},\frac{d(d-1)}{M_\alpha(M_\alpha-1)}\right\}.
\end{equation}
Obviously, $\kappa_->0$ because $r>0$. Moreover, the upper bound implies that
\begin{equation}
\bigforall_\alpha\quad r\leq\frac{d}{M_\alpha}\qquad{\rm and}\qquad \bigforall_\alpha (M_\alpha-1)r\leq\frac{d}{d-1}{M_\alpha}.
\end{equation}
Taking the sum over $\alpha=1,\ldots,N$ results in the following conditions,
\begin{equation}
r\leq\min\left\{\frac{d\mu}{N},\frac{d\mu}{d+1}\right\},
\end{equation}
which is equivalent to $\kappa_+\geq\kappa_-$.

\section{Entanglement breaking channels}\label{AppG}

Let us prove the following proposition.

\begin{Proposition}
If $E_{\alpha,k}$ belong to the class of measurements from Proposition \ref{Es}, for which $x_\alpha-y_\alpha=sw_\alpha$, then it holds that
\begin{equation}\label{Eww}
\sum_{\alpha=1}^N\sum_{k=1}^{M_\alpha}\frac{1}{w_\alpha} E_{\alpha,k}\otimes
E_{\alpha,k}=\kappa_+\mathbb{I}_d\otimes\mathbb{I}_d+\kappa_-\mathbb{F}_d
\end{equation}
with
\begin{equation}\label{Kappas}
\kappa_+=\frac{N-s}{d},\qquad\kappa_-=s.
\end{equation}
\end{Proposition}

Calculations are analogical to those in Appendix \ref{AppF}. We denote the sum of entanglement breaking channels by $\widetilde{\Phi}=\sum_{\alpha=1}^N\widetilde{\Phi}_\alpha$ and compute
\begin{equation}
\Tr(\widetilde{\Phi}[X]E_{\beta,\ell})
=\frac{x_\beta-y_\beta}{w_\beta}\Tr(E_{\beta,\ell}X)
+\frac 1d (Nw_\beta-x_\beta+y_\beta)\Tr X.
\end{equation}
This equation is then rewritten into
\begin{equation}\label{Trace}
\Tr\left\{\left(\widetilde{\Phi}[X]-
\Big[\frac{x_\beta-y_\beta}{w_\beta}X+\frac{Nw_\beta-x_\beta+y_\beta}{dw_\beta}\mathbb{I}_d\Tr(X)\Big]
\right)E_{\beta,\ell}\right\}=0,
\end{equation}
which holds for any $E_{\beta,\ell}$ and Hermitian operator $X$. The terms that multiply $\Tr X$ and $\Tr E_{\beta,\ell}X$ are constant in $\beta$ and $\ell$ if $x_\alpha-y_\alpha=sw_\alpha$. Then, 
\begin{equation}
\widetilde{\Phi}=s\oper+(N-s)\Phi_0=\kappa_-\oper+d\kappa_+\Phi_0,
\end{equation}
where $\kappa_{\pm}$ are given by eq. (\ref{Kappas}). By the Choi-Jamio{\l}kowski isomorphism, we arrive at the formula in eq. (\ref{Eww}).

\section{Dual frame operators}\label{AppH}

Let us begin the proof by assuming that every frame operator
\begin{equation}
F_{\alpha,k}=a_\alpha E_{\alpha,k}-b_\alpha\mathbb{I}_d
\end{equation}
dual to a given $E_{\alpha,k}$ is linearly proportional only to $E_{\alpha,k}$ and the identity operator. In this case,
\begin{equation}
p_{\beta,\ell}=\Tr(\rho E_{\beta,\ell})=\sum_{\alpha=1}^N\sum_{k=1}^{M_\alpha}
p_{\alpha,k}\Big[a_\alpha\Tr(E_{\alpha,k}E_{\beta,\ell})-b_\alpha w_\beta\Big].
\end{equation}
Using the properties of the measurement operators and the fact that $\{p_{\alpha,k};\,k=1,\ldots,M_\alpha\}$ is a probability distribution, we arrive at
\begin{equation}
p_{\beta,\ell}=p_{\beta,\ell}a_\beta(x_\beta-y_\beta)+a_\beta(y_\beta-z_{\beta\beta})
+\sum_{\alpha=1}^Na_\alpha z_{\alpha\beta}-\sum_{\alpha=1}^Nb_\alpha w_\beta,
\end{equation}
where $z_{\beta\beta}=d/M_\beta^2$. This equality must hold for all $p_{\beta,\ell}$, which means that
\begin{equation}\label{eqq}
a_\beta=\frac{1}{x_\beta-y_\beta},\qquad 
a_\beta(y_\beta-z_{\beta\beta})=\sum_{\alpha=1}^N\left(b_\alpha w_\beta-a_\alpha z_{\alpha\beta}\right).
\end{equation}
Now, observe that
\begin{equation}
a_\beta(y_\beta-z_{\beta\beta})=\frac{y_\beta-z_{\beta\beta}}{x_\beta-y_\beta}
=-\frac{w_\beta}{d},\qquad a_\alpha z_{\alpha\beta}=\frac{w_\beta}{d(x_\alpha-y_\alpha)}.
\end{equation}
Therefore, the terms proportional to $w_\beta$ in the second formula from eq. (\ref{eqq}) cancel out. The solution for $a_\alpha$ and $b_\alpha$ reads
\begin{equation}
a_\alpha=\frac{1}{x_\alpha-y_\alpha},\qquad b_\alpha=\frac{w_\alpha}{d(x_\alpha-y_\alpha)}-\frac{1}{dN}.
\end{equation}

\section{Index of Coincidence}\label{AppI}

Recall that $\rho$ can be written in terms of the dual frame $F_{\alpha,k}$ as
\begin{equation}
\rho=\sum_{\alpha=1}^N\sum_{k=1}^{M_\alpha}p_{\alpha,k}F_{\alpha,k},
\end{equation}
where $\{p_{\alpha,k};\,k=1,\ldots,M_\alpha\}$ are $N$ probability distributions.
To calculate $\Tr \rho^2$, we first show that
\begin{equation}
\Tr(F_{\alpha,k}F_{\beta,\ell})=\frac{1}{(x_\alpha-y_\alpha)(x_\beta-y_\beta)}
[\Tr(E_{\alpha,k}E_{\beta,\ell})-z_{\alpha\beta}]+\frac{1}{dN^2},
\end{equation}
where
\begin{equation}
\Tr(E_{\alpha,k}E_{\beta,\ell})=\delta_{\alpha\beta}\delta_{k\ell}(x_\alpha-y_\alpha)
+\delta_{\alpha\beta}(y_\alpha-z_{\alpha\beta})+z_{\alpha\beta}.
\end{equation}
Using the above equations, it is straightforward to calculate
\begin{equation}\label{rho2}
\Tr\rho^2=\frac 1d +\sum_{\alpha=1}^N\frac{1}{x_\alpha-y_\alpha}\left(\sum_{k=1}^{M_\alpha}p_{\alpha,k}^2
-\frac{1}{M_\alpha}\right).
\end{equation}
Observe that the index of coincidence $C$ does not appear in the formula for $\Tr\rho^2$ unless $x_\alpha-y_\alpha\equiv r$. 
Therefore, in order to find the relation between $\Tr\rho^2$ and $C$, one needs a constant $x_\alpha-y_\alpha\equiv r$. Only under this assumption, eq. (\ref{rho2}) simplifies to
\begin{equation}
\Tr\rho^2=\frac{C-\mu}{r}+\frac 1d,\qquad 
\mu=\sum_{\alpha=1}^N\frac{1}{M_\alpha}.
\end{equation}
Finally, due to $\Tr\rho^2\leq 1$, the index of coincidence is bounded by
\begin{equation}
C_{\max}=\frac{d-1}{d}r+\mu
\end{equation}
with the upper bound reached by pure states.

\bibliography{C:/Users/cyndaquilka/OneDrive/Fizyka/bibliography}

\begin{thebibliography}{10}
\providecommand{\url}[1]{\texttt{#1}}
\providecommand{\urlprefix}{URL }
\providecommand{\eprint}[2][]{\url{#2}}

\bibitem{Zhou}
P.~Zhou, J. Phys. A: Math. Theor. \textbf{45}, 215305 (2012).

\bibitem{Song}
J.-F. Song and Z.-Y. Wang, Int. J. Theor. Phys. \textbf{50}, 2410 (2011).

\bibitem{Hirsch}
F.~Hirsch, M.~T. Quintino, J.~Bowles, and N.~Brunner, Phys. Rev. Lett.
  \textbf{111}, 160402 (2013).

\bibitem{Blume}
R.~Blume-Kohout, J.~O.~S. Yin, and S.~J. van Enk, Phys. Rev. Lett.
  \textbf{105}, 170501 (2010).

\bibitem{Siendong}
S.~Huang, Phys. Lett. A \textbf{377}, 448 (2013).

\bibitem{Prugovecki}
E.~Prugove{\v{c}}ki, Int. J. Theor. Phys. \textbf{16}, 321--331 (1977).

\bibitem{Renes}
J.~M. Renes, R.~Blume-Kohout, A.~J. Scott, and C.~M. Caves, J. Math. Phys.
  \textbf{45}, 2171 (2004).

\bibitem{Schwinger}
J.~Schwinger, Proc. Nat. Acad. Sci. U.S.A. \textbf{46}, 570 (1960).

\bibitem{Szarek}
M.~B. Ruskai, S.~Szarek, and E.~Werner, Linear Algebra Appl. \textbf{347(1-3)},
  159--187 (2002).

\bibitem{Kalev}
A.~Kalev and G.~Gour, New J. Phys. \textbf{16}, 053038 (2014).

\bibitem{Gour}
A.~Kalev and G.~Gour, J. Phys. A: Math. Theor. \textbf{47}, 335302 (2014).

\bibitem{Ivanovic}
I.~D. Ivanovi{\'{c}}, J. Phys. A: Math. Theor. \textbf{14}, 3241 (1981).

\bibitem{Dieks}
D.~Dieks, Phys. Lett. A \textbf{126}, 303 (1988).

\bibitem{semi-SIC}
I.~J. Geng, K.~Golubeva, and G.~Gour, Phys. Rev. Lett. \textbf{126}, 100401
  (2021).

\bibitem{EOM22}
L.~Feng and S.~Luo, Phys. Lett. A \textbf{445}, 128243 (2022).

\bibitem{EOM24}
L.~Feng, S.~Luo, Y.~Zhao, and Z.~Guo, Phys. Rev. A \textbf{109}, 012218 (2024).

\bibitem{EOMq3}
Y.~Zhao, Z.~Guo, L.~Feng, S.~Luo, and T.-L. Lee, Phys. Lett. A \textbf{495},
  129314 (2024).

\bibitem{SIC-MUB}
K.~Siudzi\'{n}ska, Phys. Rev. A \textbf{105}, 042209 (2022).

\bibitem{SICMUB_entropic}
F.~Huang, L.~Tang, and M.-Q. Bai, Int. J. Theor. Phys. \textbf{62}, 126 (2023).

\bibitem{SICMUB_entropic2}
F.~Huang, F.~Wu, L.~Tang, Z.-W. Mo, and M.-Q. Bai, Phys. Scr. \textbf{98},
  105103 (2023).

\bibitem{SICMUB_App3}
L.~Tang and F.~Wu, Quantum Inf. Process. \textbf{22}, 65 (2023).

\bibitem{Alber2}
M.~Schumacher and G.~Alber, Phys. Scr. \textbf{98}, 115234 (2023).

\bibitem{SICMUB_BZ}
L.~Tang, F.~Wu, Z.~Mo, and M.~Bai, Phys. Scr. \textbf{98}, 125225 (2023).

\bibitem{SICMUB_design}
F.~Huang, F.~Wu, L.~Tang, Z.-W. Mo, and M.-Q. Bai, Phys. Scr. \textbf{98},
  105103 (2023).

\bibitem{SICMUB_App}
L.~Tang and F.~Wu, Phys. Scr. \textbf{98}, 065114 (2023).

\bibitem{SICMUB_App2}
L.~Tang, Quantum Inf. Process. \textbf{22}, 57 (2023).

\bibitem{Lai}
L.~Lai and S.~Luo, Commun. Theor. Phys. \textbf{75}, 065101 (2022).

\bibitem{SICMUB_App4}
L.~Tang and F.~Wu, Results Phys. \textbf{51}, 106663 (2023).

\bibitem{Alber}
M.~Schumacher and G.~Alber, Phys. Rev. A \textbf{108}, 042424 (2023).

\bibitem{SICMUB_Pmaps}
K.~Siudzi\'{n}ska, Sci. Rep. \textbf{12}, 10785 (2022).

\bibitem{Alber3}
M.~Schumacher and G.~Alber, \textit{Conditions for the existence of positive
  operator valued measures} (2023), arXiv:2310.12302 [quant-ph].

\bibitem{Adamson}
R.~B.~A. Adamson and A.~M. Steinberg, Phys. Rev. Lett. \textbf{105}, 030406
  (2010).

\bibitem{Scott}
A.~J. Scott, J. Phys. A: Math. Gen. \textbf{39}, 13507 (2006).

\bibitem{Renes2}
J.~M. Renes, Phys. Rev. A \textbf{70}, 052314 (2004).

\bibitem{Cr1}
N.~J. Cerf, M.~Bourennane, A.~Karlsson, and N.~Gisin, Phys. Rev. Lett.
  \textbf{88}, 127902 (2002).

\bibitem{Spengler}
C.~Spengler, M.~Huber, S.~Brierley, T.~Adaktylos, and B.~C. Hiesmayr, Phys.
  Rev. A \textbf{86}, 022311 (2012).

\bibitem{ESIC}
J.~Shang, A.~Asadian, H.~Zhu, and O.~G{\"{u}}hne, Phys. Rev. A \textbf{98},
  022309 (2018).

\bibitem{Graydon}
M.~A. Graydon and D.~M. Appleby, J. Phys. A: Math. Theor. \textbf{49}, 085301
  (2016).

\bibitem{Graydon2}
M.~A. Graydon and D.~M. Appleby, J. Phys. A: Math. Theor. \textbf{49}, 33LT02
  (2016).

\bibitem{Wang}
K.~Wang, N.~Wu, and F.~Song, Phys. Rev. A \textbf{98}, 032329 (2018).

\bibitem{SICMUB_channels}
K.~Siudzi\'{n}ska, \textit{Non-Markovian quantum dynamics from symmetric
  measurements} (2024), arXiv:2402.04415 [quant-ph].

\bibitem{Rastegin5}
A.~E. Rastegin, Eur. Phys. J. D \textbf{67}, 269 (2013).

\bibitem{Koashi}
M.~Koashi, New J. Phys. \textbf{11}, 045018 (2009).

\bibitem{Coles}
P.~J. Coles, M.~Berta, M.~Tomamichel, and S.~Wehner, Rev. Mod. Phys.
  \textbf{89}, 015002 (2017).

\bibitem{Guhne2}
O.~G{\"u}hne, Phys. Rev. Lett. \textbf{92}, 117903 (2004).

\bibitem{Rastegin4}
A.~E. Rastegin, Quant. Inf. Proc. \textbf{15}, 2621--2638 (2016).

\bibitem{AEUR}
R.~Adamczak, R.~Lata{\l}a, Z.~Pucha{\l}a, and K.~\.{Z}yczkowski, J. Math. Phys.
  \textbf{57}, 032204 (2016).

\bibitem{MEUR}
Z.~Pucha{\l}a, {\L}.~Rudnicki, and K.~\.{Z}yczkowski, J. Phys. A: Math. Theor.
  \textbf{46}, 272002 (2013).

\bibitem{Sanchez}
J.~S{\'{a}}nchez, Phys. Lett. A \textbf{173}, 233--239 (1993).

\bibitem{Maassen}
H.~Maassen and J.~B.~M. Uffink, Phys. Rev. Lett. \textbf{60}, 1103 (1988).

\bibitem{Renyi}
A.~R\'{e}nyi, \textit{On measures of entropy and information}, in
  \textit{Proceedings of 4th Berkeley symposium on mathematical statistics and
  probability, Vol. I, 547-561},  University of California Press, Berkeley
  1961.

\bibitem{Raus}
R.~Raussendorf and H.~J. Briegel, Phys. Rev. Lett. \textbf{86}, 5188 (2001).

\bibitem{Piveteau}
A.~Piveteau, J.~Pauwels, E.~Hakansson, S.~Muhammad, M.~Bourennane, and
  A.~Tavakoli, Nature Commun. \textbf{13}, 7878 (2022).

\bibitem{Ekert}
A.~K. Ekert, Phys. Rev. Lett. \textbf{67}, 661 (1991).

\bibitem{Brassard}
C.~H. Bennett, G.~Brassard, C.~Cr{\'e}peau, R.~Jozsa, A.~Peres, and W.~K.
  Wootters, Phys. Rev. Lett. \textbf{70}, 1895 (1993).

\end{thebibliography}
\bibliographystyle{C:/Users/cyndaquilka/OneDrive/Fizyka/beztytulow2}

\end{document}